# Are nationally oriented journals indexed in Scopus becoming more international? The effect of publication language and access modality

Henk F. Moed [1], Felix de Moya-Anegon [2], Vicente Guerrero-Bote[3], Carmen Lopez-Illescas [4]

[1] *henk.moed@uniroma1.it*
Sapienza University of Rome, Italy

[2]*felix.moya@scimago.es*
SCImago Group, Madrid, Spain

[3]*guerrero@unex.es*
SCImago Group, Dept. Information and Communication, University of Extremadura, Badajoz, Spain

[4]*carmlopz@gmail.com*
University Complutense of Madrid. Information Science Faculty. Dept. Information and Library Science, SCImago Group, Spain

## Abstract

An exploratory-descriptive analysis is presented of the national orientation of scientific-scholarly journals as reflected in the affiliations of publishing or citing authors. It calculates for journals covered in Scopus an Index of National Orientation (INO), and analyses the distribution of INO values across disciplines and countries, and the correlation between INO values and journal impact factors. The study did *not* find solid evidence that journal impact factors are good measures of journal internationality in terms of the geographical distribution of publishing or citing authors, as the relationship between a journal's national orientation and its citation impact is found to be inverse U-shaped. In addition, journals publishing in English are not necessarily internationally oriented in terms of the affiliations of publishing or citing authors; in social sciences and humanities also USA has their nationally oriented literatures. The paper examines the extent to which nationally oriented journals entering Scopus in earlier years, have become in recent years more international. It is found that in the study set about 40 per cent of such journals does reveal traces of internationalization, while the use of English as publication language and an Open Access (OA) status are important determinants.

## 1. Introduction

Journal internationality is an important aspect, both for researchers selecting the journals to read or publish in, as well as for research managers and policy makers to assess research activities and evaluate funding policies. In many assessment processes at the institutional or national level, publishing in "international" journals is considered a valid criterion in the evaluation of research output of individual researchers, groups and institutions. The use of journal impact factors (JIFs) and related citation based indicators of journal impact seems to be based on the assumption that JIFs are good measures of journal internationality. But what is the empirical evidence supporting this assumption?

In the perception of many assessors of scientific journal performance, the notion of "international journal" has two connotations. The first relates to *journal quality*, and is synonymous with the expressions such as "of international quality", or "among the best journals in the field". The second connotation refers to the *geographical* distribution of the authors publishing in a journal or citing the journal. In the latter case, "international" means "used by authors from all over the world". Conversely, a national journal is defined as a journal in which the major part of papers is (co-) authored by – or cited by – researchers from one single country.

The current paper focuses on the *geographical* dimension of (inter)national orientation. There are many ways to construct bibliometric measures of a journal's international or national orientation (Zitt and Bassecoulard, 1998). Moed (2005) introduced an Index of National Orientation (INO), defined as "the share of the papers from the country most frequently publishing in a journal, relative to the total number of papers published in the journal. A purely national journal would have an INO value of 100 per cent (Moed, 2005, pp.131-132)." In the current paper this INO concept is extended.

The current paper consists of three parts. The *first* part is *exploratory-descriptive*. It computes in Section 2 two indices of a journal's national orientation, one based on affiliation countries of *publishing* authors, and a second on the affiliations of authors *citing* a journal, and denoted as INO-P and INO-C, respectively. It presents the distribution of INO values of journals across disciplines and countries, and compares the INO distribution based on Scopus journals with one derived in an earlier study from the ISI citation indexes (currently Clarivate's Web of Science). It gives special attention to *social sciences and humanities*, disciplines that are often studied in recent bibliometric research (e.g., Kulczycki et al., 2018; Kulczycki, Rozkosz & Drabek, 2019; Bocanegra-Valle, 2019).

The *second* part analyses in Section 3 the statistical relationship between INO and two other measures related to journal quality and internationality, namely the citation impact measured by a journal impact factor, and also the percentage of internationally co-authored papers. It is in this part that the above mentioned assumption that journal impact factors are good measures of journal internationality is being tested.

The effect of being indexed in Scopus upon the development of a journal's international orientation over time is examined in the *third* part of the paper. During the past years, several studies have been published of the effects of indexing journals in publication or literature databases upon their visibility and geographical orientation (e.g., Ainsworth & Russell, 2018; Bucher, 2018; Toth, 2018; Macan, Pikic, A., & Mayer, 2012). Other studies claimed that to acquire global visibility and impact, it is sufficient to have one's paper included in an international database.

- According to Reedijk & Moed (2009), the impact factor value of a journal is becoming increasingly less important for authors and readers. They argued that electronic publishing reduces the importance of the journal impact factor as criterion to purchase or read a journal, as the contents of large numbers of journals are available in or via large electronic literature databases such as Web of Science or Scopus, and libraries increasingly purchase a complete package of journals from a publisher electronically, with a smaller number of printed copies.
- A citation study by Acharya et al. (2014) reports evidence that the fraction of top-cited articles published in non-elite journals increased steadily over 1995-2013, and concludes that "now that finding and reading relevant articles in non-elite journals is about as easy as finding and reading articles in elite journals, researchers are increasingly building on and citing work published everywhere".

Section 4 of the current paper address the following research questions: When nationally oriented journals start being covered by the large scientific literature database Scopus, how does their national orientation develop over time? Do such journals become more internationally oriented? Or do they remain as nationally oriented as they were when they entered the database? What is the statistical effect of a journal's *publication language* and the *access modality* ("OA versus non-OA") upon the trend in its national geographical orientation?

## 2. Indicators of a journal's national geographical orientation

*Data collection*

For each source journal indexed in Scopus, data were extracted on the number of publications and citations by country of the publishing or citing authors, and for the time period 1996-2017. In a preliminary step, publication years were grouped into overlapping three-year time periods, and the total number of publications was calculated during each 3-year time period. In the current study, only these aggregated 3-year counts were available. They constitute the denominator in the calculation of a journal impact factor applying a three-year window, counting, for instance, citations in 2017 to publications from 2014-2016. The first and last citation year for which data were available were 1999 and 2017, respectively. Scopus indexes not only journals but also conference proceedings and books, and in principle all article types except meeting abstracts and book reviews. The current study analysed articles, reviews, notes and mini reviews published in journals. These types are labelled as "papers" throughout the current article.

To eliminate journals with discontinuous coverage in Scopus over the years, those being "inactive" in the most recent 3-year period, indexed only in a few years, or containing only very few papers, it was decided to take into account only journals active in 2017, with papers in at least four subsequent 3-year periods, and publishing papers in each year between its starting year in Scopus and 2017, with on average at least 20 papers per 3-year period. There are about 23,000 of such journals. Section 2 analyses this total set, whole Section 3 is based on the subset of about 8,000 journals that entered Scopus between 1997 and 2012.

*Two indices of national geographic orientation*

The measurement of the degree of journal (inter)nationality explored in this paper is based on the affiliation countries of authors *publishing* articles in the journal, or those of the authors *citing* it. Such countries are labelled as "publishing or citing *author countries*" throughout this paper. A *first* indicator of national geographical orientation of a journal relates to the authors publishing in the journal, and is defined as the percentage of papers (co-) authored by researchers from the country *publishing* the largest number of papers to the journal (INO-P). For instance, if a journal has a INO-P value of 80 percent, this means that there is one country that accounts for 80 per cent of all papers published in that journal. It must me noted that a certain fraction of these papers can be expected to have authors from other countries as well, and thus reflect international co-authorship. A *second* indicator of national geographical orientation relates to the authors *citing* a particular journal (INO-C), and is defined as the percentage of citations given by researchers from the country contributing the largest number of citations to the journal.

If one defines a national journal as a journal in which the number of papers published from the most productive country exceeds a certain threshold, the percentage of thus-defined national journals naturally depends upon the threshold value. This is clearly illustrated in Figure 1. It presents for all about 23,000 journals active in 2017 in Elsevier's Scopus the percentage of "national" journals as a function of the applied "nationality" threshold. Figure 1 shows for instance that in 45 per cent of journals one country accounts for more than 50 per cent of published papers in a journal (possibly partly in collaboration with other countries), while in 9 percent of journals one country publishes more than 90 per cent of all papers. The relationship between percentage of national journals and "nationality threshold" is slightly convex and does for threshold values of 10 % or higher not deviate strongly from linearity.

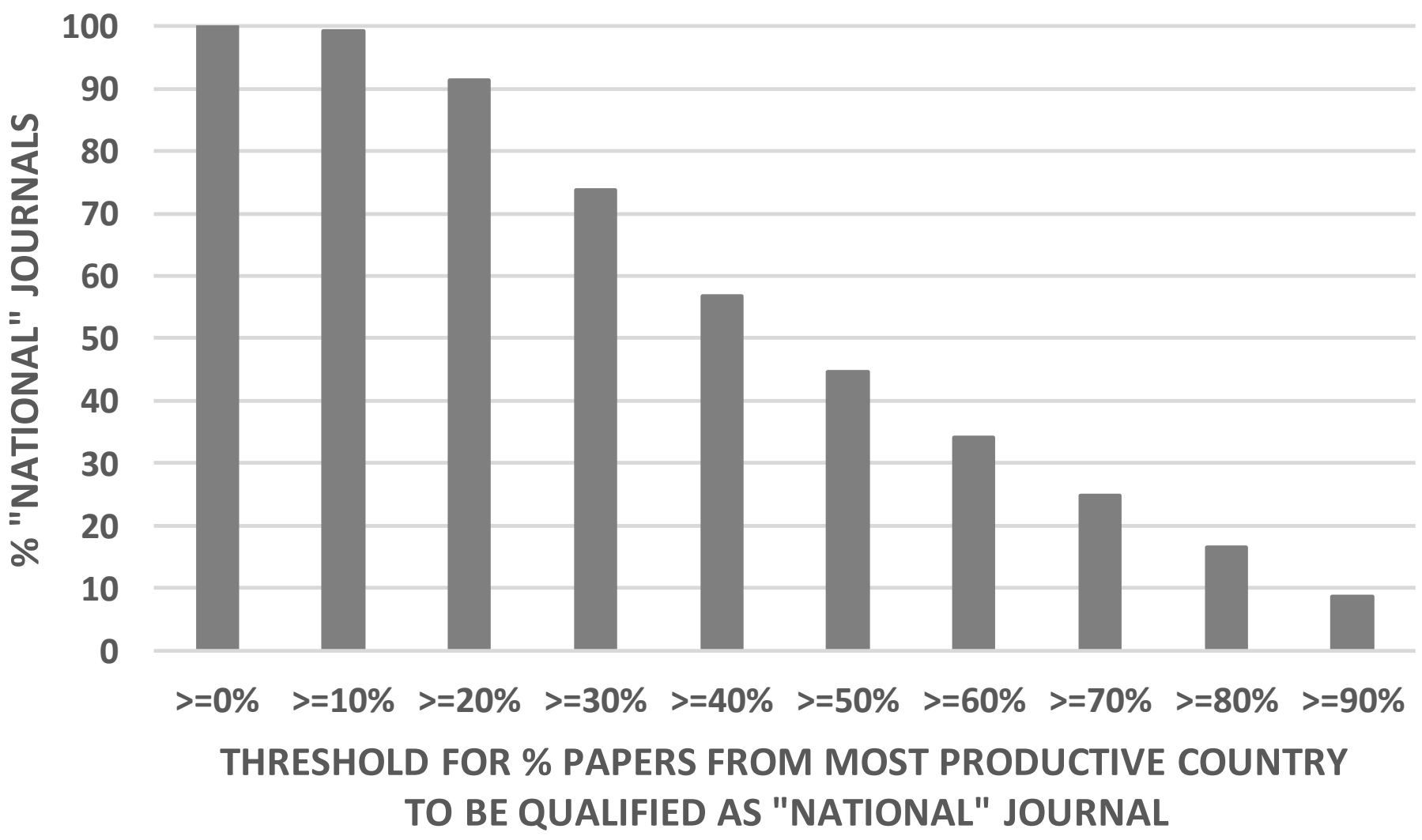

Figure 1. The relationship between percentage of national journals and "nationality threshold". It shows for instance that for 24 per cent of journals the most productive country accounts for at least 70 per cent of all papers.

*Differences between INO-P and INO-C*

Figure 2 presents the distribution of INO-P and INO-C values among journals. It shows substantial differences between the publication- and the citation-based distribution. The citation distribution is more concentrated in the lower classes than its publication-based counterpart. This means that the journals' citation impact tends to be geographically *broader* than their publication packets.

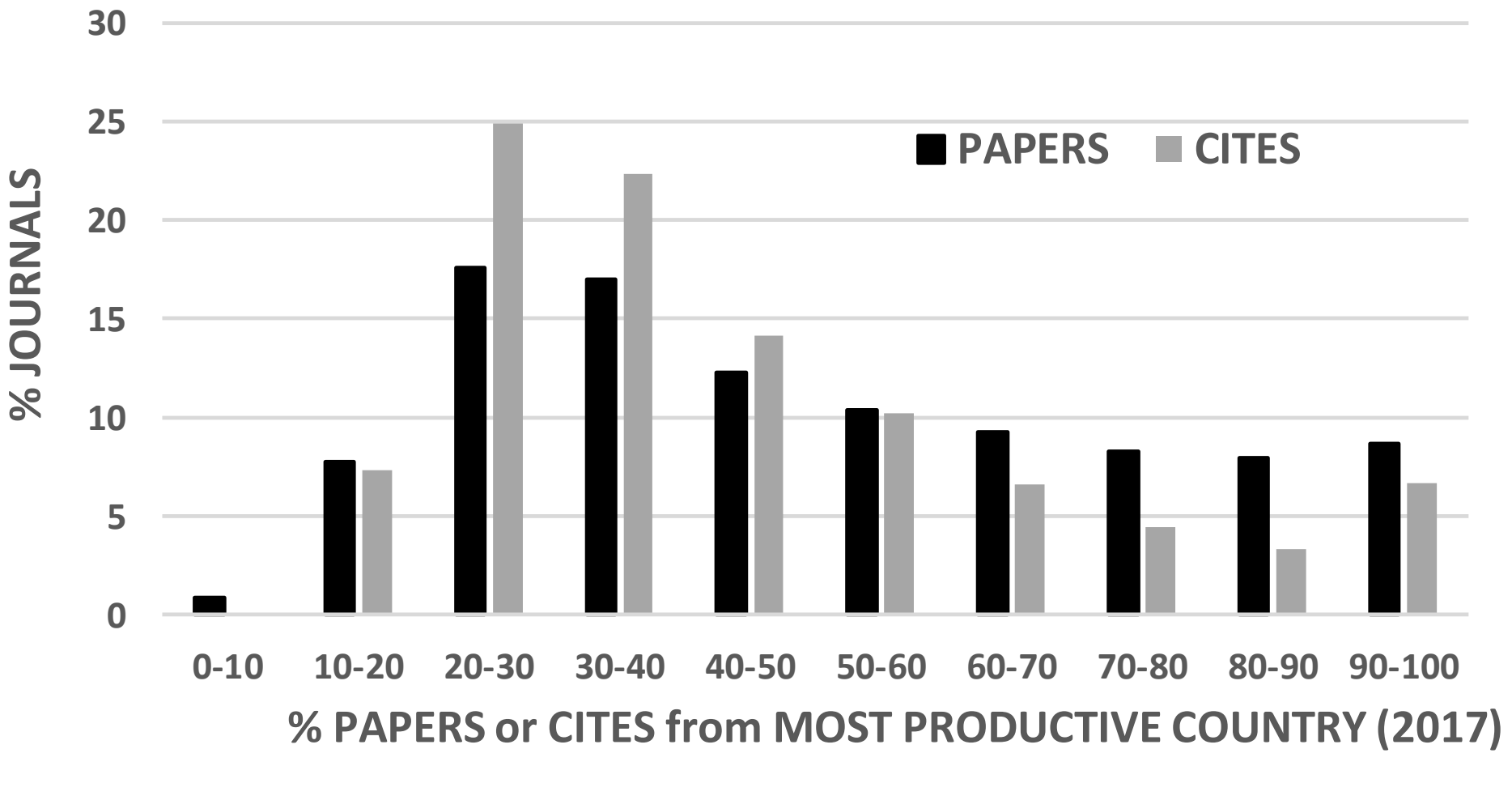

Figure 2. Distribution across journals of the percentage of papers and citations by authors from the most productive country (INO-P and INO-C, respectively).

*Differences among research fields*

A starting point is a classification of journals into 27 disciplines in Scopus. These were grouped into 5 main disciplines. Since *social sciences and humanities* subject fields in this classification contain journals from *natural and life sciences* that only *occasionally* publish papers on social science or humanities subjects (for instance, all papers in the journals *Science* and *PNAS* are assigned to at least one of these disciplines), only journals *exclusively* linked to *social sciences and/or humanities* and not to any other discipline were included in a main discipline *social sciences and humanities,* denoted as *SOCHUM.* Other main disciplines are: *biomedical research* (BIOMED); *clinical medicine* (CLIMED); *engineering* (ENGIN); and *natural sciences* (NATSCI).

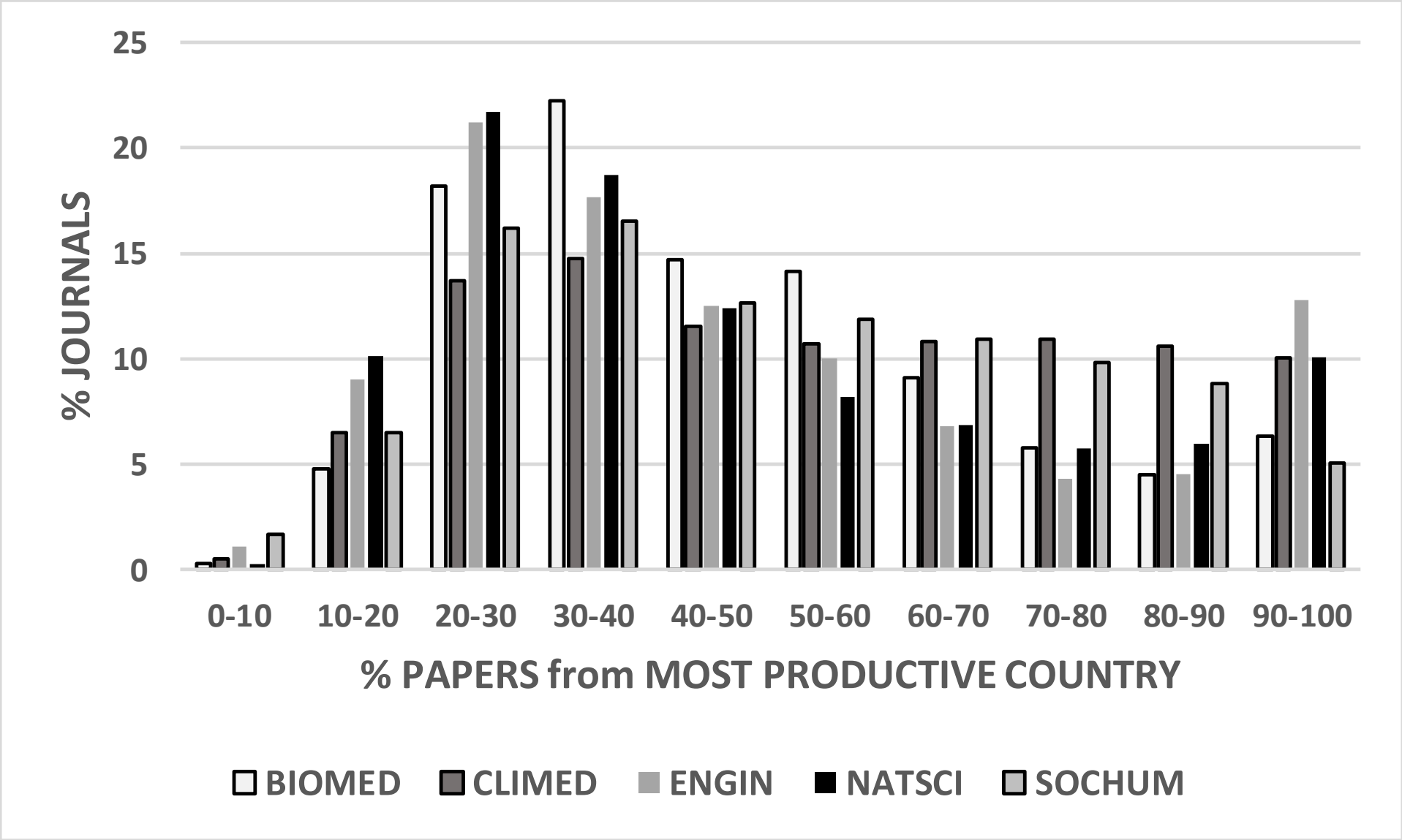

Figure 3. The distribution across journals and per main discipline of the percentage of papers published by authors from the most productive country (INO-P). Focusing on the class of journals in which INO-P exceeds 90 per cent, the figure shows that engineering has the largest share of journals in this class (13 %), and *social sciences and humanities* the smallest (around 5 %).

Figure 3 shows that the distribution for *natural sciences* shows the largest similarity with that for all fields combined presented in Figure 1. This is not surprising, as this field contributes by far the largest number of journals. In all disciplines the largest percentage of journals can be found in INO classes 20-30% or 30-40%. But in the 70-100% range there are substantial differences among disciplines. The percentage of journals in *clinical medicine* hardly shows a decline. While *natural sciences*, *biomedical research* and *engineering* show an increase in the 90-100% range compared to the 80-90% range, *social sciences and humanities* reveal a decline.

*Differences among countries*

Figure 4 presents for major countries the percentage of articles in journals for which INO-P exceeds 80 per cent. It includes only countries with more than *1,000 publications per year* during 2014-2016. In the set of countries of which *more than 20 per cent* of papers is published in national journals with INO-P>80, the dominant group is that of 1O Central and East European countries (Russia, Ukraine and Turkey, Bulgaria, Poland, Romania, Serbia, Hungary, Croatia and Slovenia), followed by 4 Latin American nations (Cuba, Brazil, Mexico and Colombia), two Asian countries (China and Pakistan), and Spain and South Africa.

Countries with *less than 10 per cent* of papers in journals with INO-P>80 include 6 Anglo-Saxon countries (USA, Canada, Australia, Great Britain, Hong Kong and Singapore); 12 European countries (France, Germany, Italy, Austria, Switzerland, Slovakia, Netherlands, Norway, Finland, Sweden, Denmark, Greece); 5 mostly Northern African countries (Egypt, Tunisia, Algeria, Morocco, and Nigeria); and 6 Asian or Middle East countries (Taiwan, Kazakhstan, Indonesia, Bangladesh, Israel and Saudi Arabia). The figure gives an impression of differences in publication practices among countries, but also of the geographical focus of the Scopus business model developed by Elsevier.

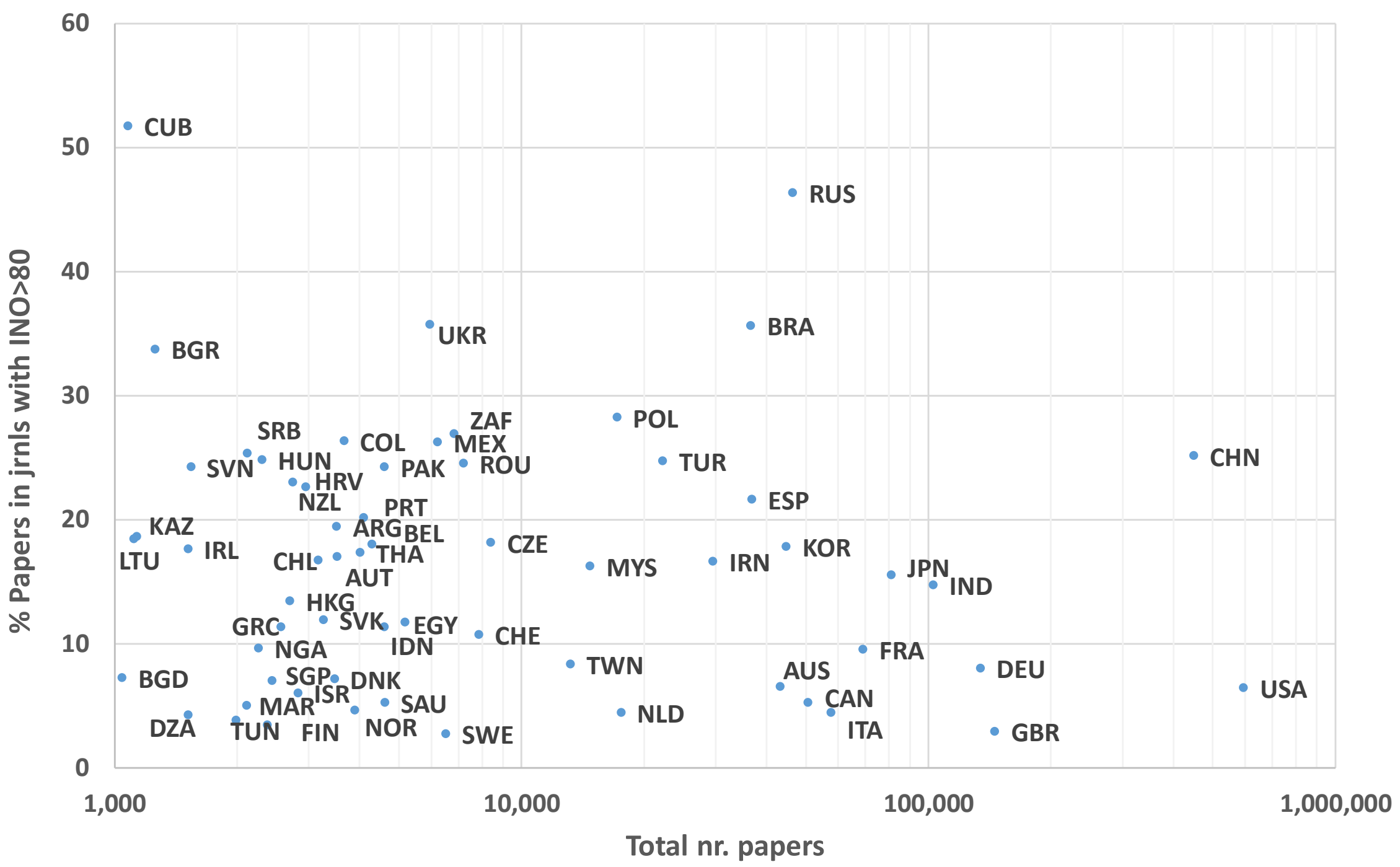

.

Figure 4. Percentage per country of articles published in nationally oriented journals. For instance, the number of papers published per year by authors affiliated with Russian institutions is between 40 and 50 thousand, while the share of these papers with INO > 80% is around 46 per cent.

Large differences exist between disciplines. Considering again publications in journals with INO-P > 80%, the fraction of papers from a country published in these journals tends to be much higher for *clinical medicine* than it is for *natural sciences*. This outcome reflects the importance of nationally oriented journals for medical practitioners active in a country. *Social sciences and humanities* (SSH) shows an interesting pattern compared to the results for all fields combined. Table 1 gives for this discipline the percentage of publications in journals with INO-P>80% for the 10 countries with the largest number of papers. For USA this percentage *for SSH* amounts to 18, apart from that of Spain the highest score of the 10 countries in this table, and much higher than the level of 6.5 per cent obtained by USA *for all fields combined*, as reflected in Figure 4 above. The same is true for France, Germany and Italy. This outcome shows that the importance of nationally oriented journals in SSH fields, publishing predominantly in the national language, is not only reflected in the publication practices of researchers in large non-Anglo-Saxon, European countries, but also in the practices of US scholars. In other words, publishing in nationally oriented journals seems to be a characteristic of the field SSH as a whole, and carried out by scholars from many countries, Anglo-Saxon or otherwise.

Table 1. Percentage of papers in nationally oriented SSH journals from 10 major countries

| Country | Number of papers per year during 2014-2016 | % Papers in journals with INO-P>80% |
|---|---|---|
| Spain | 7,153 | 26.8 |
| USA | 72,377 | 18.0 |
| France | 6,603 | 17.6 |
| Germany | 10,840 | 13.6 |
| Italy | 4,922 | 12.0 |
| Canada | 7,990 | 8.6 |
| Australia | 9,580 | 8.3 |
| China | 5,196 | 8.1 |
| Great Britain | 24,591 | 3.7 |
| Netherlands | 4,152 | 3.5 |

*Comparing Scopus and ISI Citation Indexes (currently Clarivate's Web of Science)*

Comparing the INO-P values related to the publication year 2002 obtained by Moed (2005, p. 131) for the ISI Indexes with those found in the current study based on Scopus *for the same year*, it was found that in *science* fields, *(bio-) medicine* and *engineering* the percentages of journals with an INO value greater than 90 per cent are in the combined ISI Citation Indexes on CD-Rom similar to those obtained in the current study in Scopus for publications from the same year. But for social sciences and humanities large differences were observed: at the ISI side, for *humanities & arts* and *other social sciences*, the percentage of journals with INO-P greater than 90 was 24 and 22 per cent, respectively, while in Scopus for the main discipline *Social sciences and humanities* in that year a value 8 per cent was obtained. This outcome suggests that the journals covered in social science and humanities fields showed in the beginning of the millennium in the ISI Indexes a much stronger national orientation than those published in journals indexed in Scopus with the same publication year.

## 3. Statistical relationship between a journal's indicator of national orientation and its citation impact and percentage of internationally co-authored papers.

Figure 5 shows for all fields combined a breakdown of INO-P values into deciles, and for each decile the mean field-normalized impact value, defined as the ratio of the journal's impact factor (denoted as JIF in this paper) and the mean impact factor across all journals in a discipline. JIF is based on a three-year citation window as used in the calculation of the SJR (González-Pereira, Guerrero-Bote, & Moya-Anegón, 2010) rather than a two-year citation window as in the standard impact factor published by Clarivate Analytics. INO-P and JIF show an *inverse U-shape* relationship. It is hypothesized that the classes with INO-P values between 30 and 60 contain relatively many journals used predominantly by authors from large, scientifically developed countries such as USA, Canada, UK, Germany, whose articles tend to have a relatively high citation impact. But further research is needed to obtain a better understanding of this pattern. In any case, the inverse U-shape relationship makes the calculation of linear or rank correlation coefficients between these two variables less meaningful.

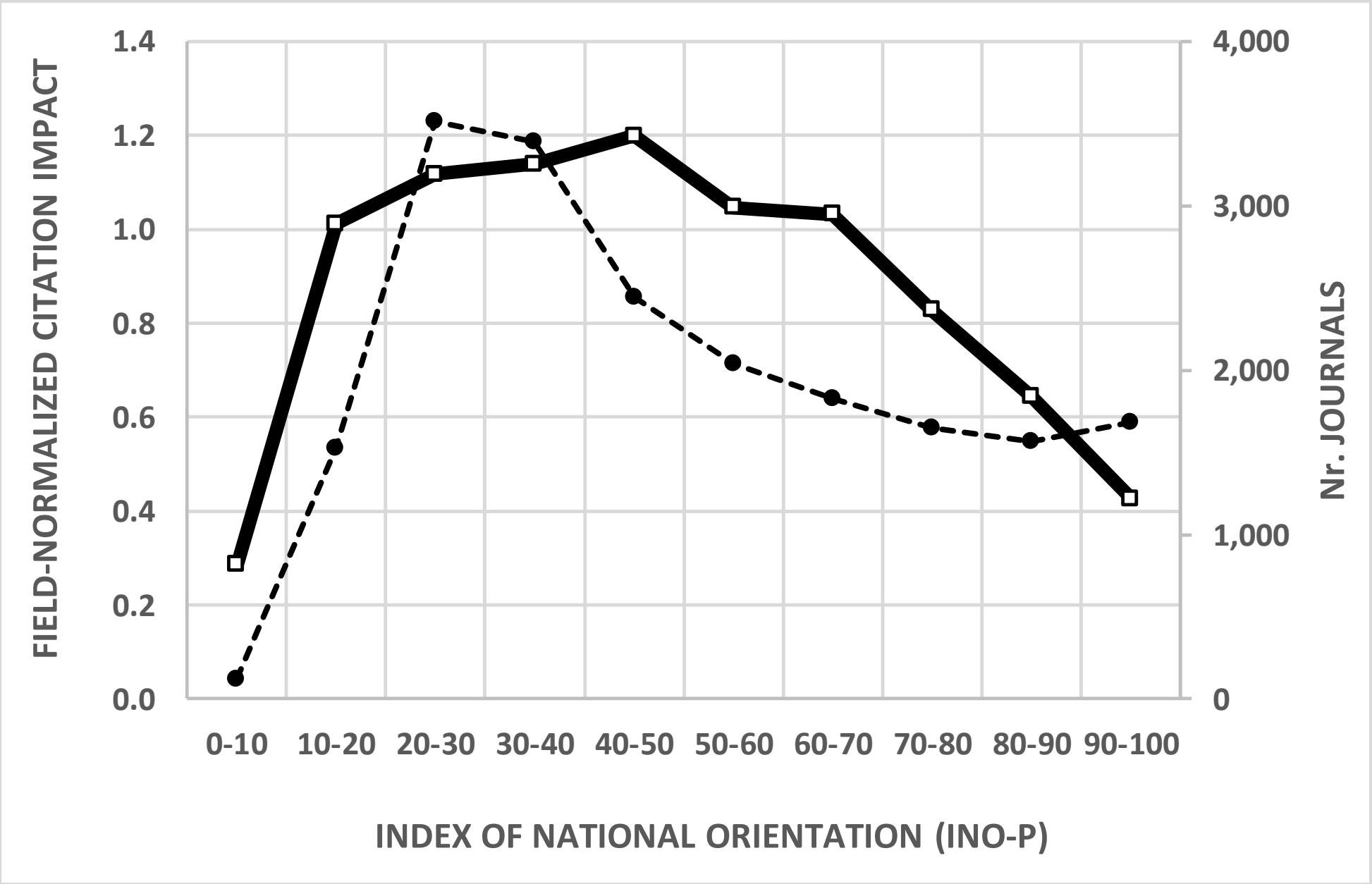

Figure 5. Field-normalized impact per INO-P decile for all fields combined. Bold line: Citation Impact; Dashed line: Number of journals.

Results per main discipline are presented in Figure 6. While journals in *natural sciences*, *clinical medicine* and *engineering* show an inverse U-shape relationship similar to the overall pattern displayed in Figure 5, the decline phase for biomedical journals starts from INO-P values above 80 per cent, and shows a steep decline. Moreover, journals from *social sciences and humanities* tend to follow a deviant pattern as they show in their decline phase only a moderately decreasing trend.

The functional relationship between the *percentage of internationally co-authored papers* and INO-P class shows a pattern similar to that for normalized impact factors versus INO-P displayed in Figure 6. For INO-P values above 40% a semi-linear decline, occurring in all main disciplines, although in *social sciences and humanities* it is more gradual than in other fields. This is at least partly due to the overall lower level of international co-authorship in this main discipline.

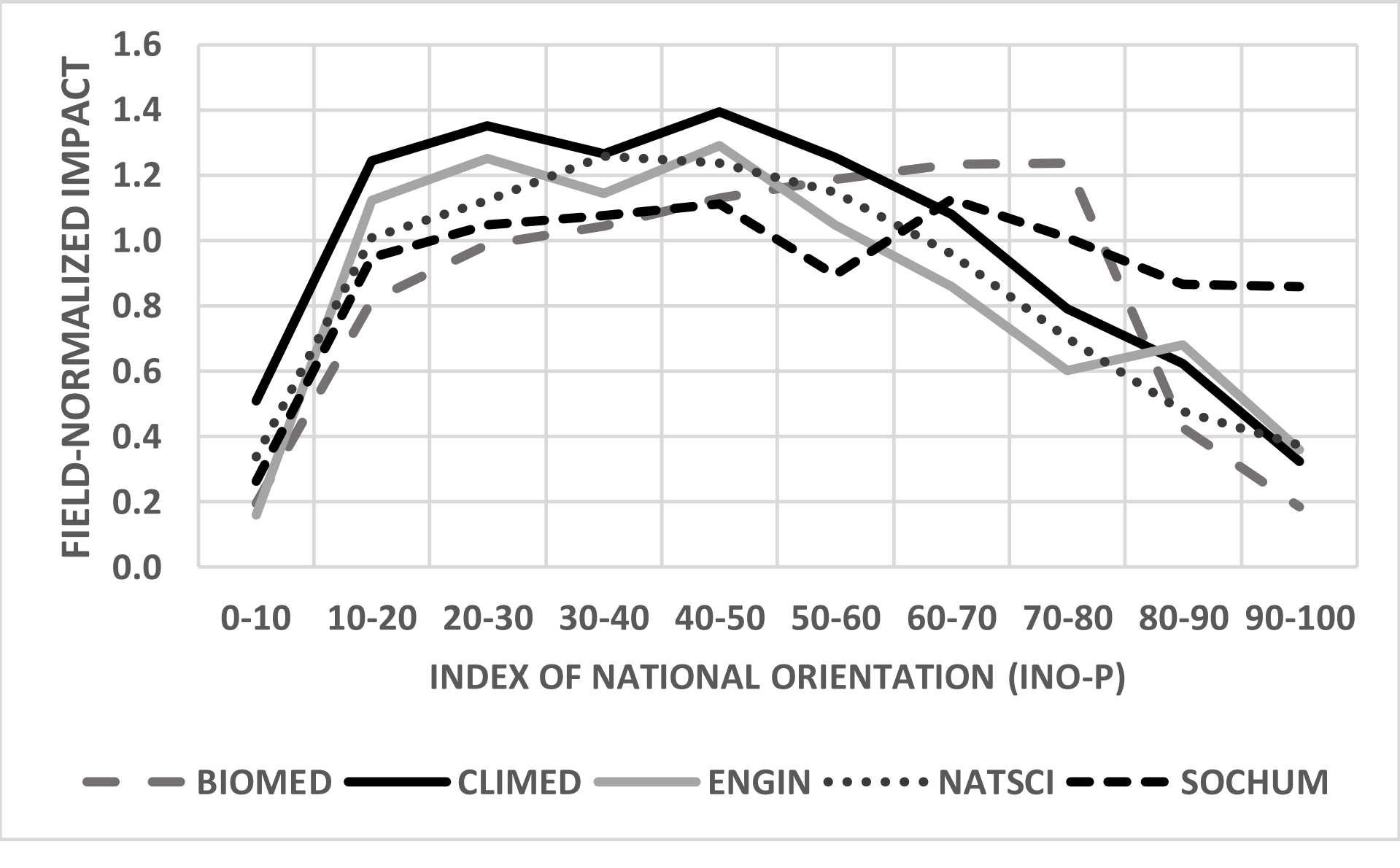

Figure 6. Field-normalized impact per INO-P decile and per main discipline

## 4. The statistical effect of indexing nationally oriented journals in Scopus upon their international orientation

In a first step, all journals were selected that entered Scopus between 1997 and 2012, and that had in their starting year in Scopus an initial INO-P value above 80 per cent. This set contained 2,192 journals. Next, for each selected journal, a growth rate was computed of the annual INO-P values over the years, based on the outcomes of a linear regression, with Initial INO-P as the dependent and the year as independent variable, and by dividing the regression coefficient by the journal's mean annual score. In addition, it was tested whether the trend in annual scores was significant or not, applying a 99 per cent confidence level.

It was found that the share of journals from the selected set showing a statistically significant decline in annual INO-P values is 39 per cent; another 5 per cent showed a significant increase, while for 56 per cent the linear regression coefficient did not differ significantly from 0. In addition, rather than analysing a trend based on linear regression, a second approach calculated for each time series a Compound Annual Growth Rate (CAGR), an indicator that totally depends upon the values at the beginning and the end of the time period considered. It was found that 82 per cent of journals revealed a negative, and 18 per cent a positive CAGR.

These outcomes are partly a statistical artefact. If one selects, for instance, from a set of objects with randomly assigned binary scores (0 or 1) in two subsequent years a sub-set of those objects that have score 1 in the first year, the probability that the score of journals in this sub-set is in the second year lower than that in the first year (i.c., score 0) is 50 per cent. Also, the significance test in the linear regression has been conducted for 2,200 cases. Even though the applied confidence level was rather high (99 per cent), in a certain fraction of cases the conclusion drawn from the test may be false. It is assumed that, although the observed percentage of nationally oriented journals (Initial INO-P>80) showing a decline in their national orientation is partly a statistical artefact, it is *not fully* determined by it. Therefore, it is of interest and meaningful to further analyse the trend data, and examine the role of *six* external factors upon the decline rate of INO-P: the journals' main discipline; the publishing country involved in their national orientation; their publication language and access modality (OA status); the year in which they entered Scopus; and the degree of national orientation in the entry year.

*The role of publication language, access modality and Scopus entry date*

Data on publication language and OA status were obtained from the Scopus Source Title list (Scopus, 2019). It is assumed that the data on publication language and OA status are valid as from the first year for which the journal was indexed in Scopus, and that these two parameters did *not* change during the time period analysed. This assumption is crucial in the interpretation of the results below.

Two analyses were conducted. A first related to the role of publication language and access modality as *separate* variables, or *in combination*. The outcomes are presented in Table 2. A second analysis applies a linear regression model with a journal's decline rate in its strongly nationally oriented journals (Initial INO-P>80%) as the dependent variable, and the publication language, OA status and the entry year in Scopus as independent variables, calculating for language and OA related measures dummy variables. Table 3 presents the main outcomes.

Table 2 clearly illustrates that journals using English as the sole publication language have a larger share of journals showing a significant decline in their INO-P than sources using, apart from English, other publication languages as well, or that use no English at all. Journals included in the DOAJ /ROAD database and labelled as OA in Table 3 show a stronger degree of internationalization than journals that are not included in this Directory (denoted as Non-OA), regardless of the publication language used, even though the largest difference between OA and non-OA journals can be observed for journals with a non-English publication language.

The outcomes of the regression analysis presented in Table 3 are consistent with those obtained in Table 2. Considering the parameter estimates, publishing solely in English is the most important factor explaining the decline in INO-P, followed by having an OA status. By contrast, as regards INO-C, OA status is the most important factor, although in this case the parameter values of the two types of factors are more similar than they are in the INO-P analysis. A journal's *entry year* in Scopus has a small, though statistically significant, negative effect. This means that nationally oriented journals entering Scopus in an early year have a stronger tendency to reveal a decline in their national orientation than such journals entered in more recent years.

Table 2. Share of journals with significant declining INO in function of publication language and OA status (p=0.01)

| Publication language | OA Status | Nr journals | % Journals with significant decline in INO-P | | % Journals with significant decline in INO-C | |
|---|---|---|---|---|---|---|
| English only | | 1,251 | 47% | + | 28% | o |
| Multiple languages including English | | 285 | 31% | - | 23% | o |
| Non-English | | 576 | 26% | -- | 15% | -- |
| | OA | 774 | 44% | o | 29% | + |
| | Non-OA | 1,418 | 36% | o | 21% | o |
| | | | | | | |
| English only | OA | 422 | 51% | + | 32% | + |
| English only | Non-OA | 829 | 44% | o | 26% | o |
| Multiple languages including English | OA | 139 | 32% | - | 26% | o |
| Multiple languages including English | Non-OA | 146 | 31% | - | 21% | o |
| Non-English | OA | 201 | 37% | o | 24% | o |
| Non-English | Non-OA | 371 | 19% | -- | 10% | -- |
| Missing language or OA status info | | 80 | | | | |
| All journals | | 2,192 | 39% | | 24% | |

Legend to Table 2: OA means: Journal included in DOAJ/ROAD. Source: Scopus (2019). If score is defined as the percentage of journals showing a significant decline with a particular language and/or OA status divided by the overall percentage of journals (39% for INO-P and 24% for INO-C, the +,-,o symbols have the following meaning: ++: score>1.5; +: 1.2<=score<=1.5; o: 0.83<score<1.2; -: 0.67=<score<=0.83; --: score<0.67.

Table 3: Results of regression analysis

<table>
<tr><th rowspan="2">Variable</th><th rowspan="2">DF</th><th colspan="2">Parameter estimate</th><th colspan="2">Parameter standard error</th><th colspan="2">t-value</th><th colspan="2">Pr>|t|</th></tr>
<tr><th>INO-P</th><th>INO-C</th><th>INO-P</th><th>INO-C</th><th>INO-P</th><th>INO-C</th><th>INO-P</th><th>INO-C</th></tr>
<tr><td>Intercept</td><td>1</td><td>400.0</td><td>515.3</td><td>55.3</td><td>79.6</td><td>+7.23</td><td>+6.47</td><td><.0001</td><td><.0001</td></tr>
<tr><td>OA</td><td>1</td><td>-1.05</td><td>-2.58</td><td>0.20</td><td>0.28</td><td>-5.32</td><td>-9.13</td><td><.0001</td><td><.0001</td></tr>
<tr><td>English only</td><td>1</td><td>-2.02</td><td>-2.22</td><td>0.26</td><td>0.37</td><td>-7.78</td><td>-5.94</td><td><.0001</td><td><.0001</td></tr>
<tr><td>Non-English</td><td>1</td><td>+0.73</td><td>1.14</td><td>0.29</td><td>0.42</td><td>+2.52</td><td>+2.72</td><td>0.0119</td><td>0.0066</td></tr>
<tr><td>Entry Year</td><td>1</td><td>-0.20</td><td>-0.26</td><td>0.03</td><td>0.04</td><td>-7.26</td><td>-6.50</td><td><.0001</td><td><.0001</td></tr>
</table>

*The effect of the degree of national orientation in the entry year (Initial INO-P)*

Conducting a regression analysis based on all 8,076 journals in the set, and adding *Initial INO-P* as independent variable, 24 per cent of journals reveals a significant decline in their national orientation. Both English as publication language and OA status are significant and negative factors, with regression parameter estimates of -2.57 and -0.53, respectively. The factor Initial INO-P value in the entry year in Scopus is significant at $p<0.001$ but relatively small: the regression parameter estimate is -0.10. The interpretation of this outcomes is unclear: it could be a statistical artefact, showing that journals with high initial INO values their entry year tend to show more often a decline than journals with lower initial INO values do.

For journals with initial INO-P values between 40 and 60 per cent, the share of journals with a significant decline was found to be 24 per cent, while 8 % show an increase and 68 % no significant trend. This sub-set of journals is interesting because there are no a-priori selection biases when assessing negative trends. Conducting a regression analysis on this sub-set similar to one presented above, it was found that publishing in English is a significant factor explaining decline of INO-P, but Open Access status is not significant at $p<0.01$. In fact, both for OA and for non-OA journals the percentage of journals with a significant decline is 24%.

*Main disciplines*

In the sub-set analysed there are appear to be no significant differences among main disciplines in the percentage of journals showing a decline. In engineering and social sciences & humanities this is 48 per cent, in *natural sciences* and *clinical medicine* 52 %, and in *biomedical research* 56 %. Including these disciplines as dummy variables in the regression analysis presented below, none of these are significant at the 99 per cent confidence level.

*Countries*

Substantial differences exist among countries. This is illustrated in Table 4. It presents for strongly nationally oriented journals with INO-P values of 80 per cent or higher the share of journals with significant declining INO-P. Journals mainly oriented in the year they entered Scopus towards USA, Japan, Brazil and Iran reveal a broadening of the citation impact compared to the overall average. China, Germany, Russia and France show a relatively weak tendency towards internationalization both in terms of authors publishing and citing authors, and Korea and Great Britain a strong tendency. Spain shows an 'average' tendency to internationalize in terms of publishing authors, but a weak tendency in terms of geographically broadening the citation impact.

Differences among countries are largely statistically explained by publication language and OA status. When including these countries via dummy variables into the regression model, and thus, from the point of view of a country, correcting for differences in in these two external factors, most countries do not show a significant decline in their national orientation (INO-P or INO-C).

Table 4. Share of journals with declining INO values by publishing country

| Publishing country | Nr. Journals with INO-P >80% in start year | % Journals with significantly declining INO up to 2017 (relative to world average) | | | |
|---|---|---|---|---|---|
| | | INO-P | | INO-C | |
| USA | 401 | 57.9 | o | 46.1 | + |
| CHN | 211 | 32.2 | -- | 24.2 | -- |
| BRA | 156 | 47.4 | o | 44.2 | + |
| IND | 141 | 55.3 | o | 36.9 | o |
| ESP | 107 | 43.0 | o | 29.0 | - |
| POL | 90 | 53.3 | o | 42.2 | o |
| JPN | 85 | 43.5 | o | 47.1 | + |
| IRN | 83 | 48.2 | o | 44.6 | + |
| DEU | 82 | 41.5 | - | 20.7 | -- |
| TUR | 80 | 43.8 | o | 37.5 | o |
| KOR | 79 | 73.4 | + | 58.2 | ++ |
| RUS | 68 | 38.2 | - | 2.9 | -- |
| GBR | 53 | 64.2 | + | 45.3 | + |
| ITA | 49 | 55.1 | o | 32.7 | o |
| FRA | 40 | 30.0 | -- | 30.0 | - |

++: score>1.5; +: 1.2<=score<=1.5; o: 0.83<score<1.2; -: 0.67=<score<=0.83; --: score<0.67. World average is defined as the overall percentage of journals with significantly declining INO-P or INO-C for the total set of 2,192 journals presented above: 50 % for INO-P and 36 % for INO-C. For instance: + in INO-P means: > 50*1.2=60%. -- in INO-C means: <34*0.67=24%. Only countries with 40 or more journals are included.

## 5. Discussion and conclusions

The results show that it does not make much sense to speak in terms of national or international journals. Therefore, it was proposed to define and calculate an indicator of national or international *orientation*. There are other measures of a journal's international orientation than those explored in the current paper. For instance, the geographical spread of a journal's editorial referee board is another, non-bibliometric one. Since the measures proposed in this paper strictly relate to geographical, publication and citation-based data, it is appropriate to express this in the acronym used to denote the indicators used, and use INO-P and INO-C rather than just INO.

There is no simple, linear relationship between INO-P or INO-C on the one hand, and two other often applied journal measures on the other, namely the field-normalized journal impact, and a journal's percentage of internationally co-authored papers. The functional relationship between the INO indicators and the latter two measures is a reversed U-shape. Calculation of linear correlation coefficients is not meaningful. Therefore, the study did *not* find solid evidence that journal impact factors are good measures of journal internationality in terms of the geographical distribution of authors or readers.

This conclusion, along with the finding that in social science and humanities (SSH) fields nationally oriented journals are as important in the USA as they are in non-English speaking countries France, Germany and Italy, has implications for the debate on the validity of bibliometric indicators. Journals that publish in English are not necessarily internationally oriented in terms of the affiliations of publishing or citing authors; also Anglo-Saxon nations may have their national literatures. It should be noted that Great Britain shows a much lower percentage of papers in national journals than USA. Possibly, UK journals tend to attract more papers by scholars from non-English speaking countries who decide to publish in English language than US journals do, and therefore have a stronger international orientation. A second factor concerns the heterogeneity of the SSH field, that includes, for instance, both econometric and philosophical journals, with different publication practices. Differences between Great Britain and USA may exist in the cognitive specialization within SSH.

It was found that 39 per cent of nationally oriented journals (Initial INO-P>80%) show an increase in their international orientation upon entering Scopus. An evaluation of this outcome depends on the perspective of the analyst. To the extent that the inclusion of nationally oriented journals in a globally marketed and distributed database is part of a policy to make these journals more international in terms of authors and impact, one should conclude that this policy has *thus far* been only partially successful. On the positive side, a substantial number of journals *does* show a positive trend in their international orientation, regardless the statistical interpretation problems highlighted in the previous section. The tendency towards internationalization is less prominent in the geographical distribution of the *citation* impact than it is if one focuses on the institutional affiliations or *publishing* authors. The fact that this ratio for citing authors is lower than that obtained for the geographical distribution of publishing authors, may reflect at least partly a time delay in article *impact* compared to *production*, but more research is needed to further investigate this hypothesis.

As regards the claim that to acquire global visibility and impact, it is sufficient to have one's paper included in an international database, the outcomes obtained in the current study do not provide equivocal evidence of its validity. They suggest that this claim is not valid across all journals entering an international database, and that publication language and access modality are important preconditions for their internationalization, at least in terms of publishing and reading authors.

*Limitations and further research*

It must be noted that the outcomes of the regression analysis do not allow to strictly separate the language and the access modality factor. Preliminary analyses suggest that not seldom newly established journals implement *both* factors at the same time. In fact, many journals making a transition from a subscription-based to an author pay based model are being fully reshaped, and change at the same time also the publication language and other editorial characteristics, including, for instance, the composition of the editorial board (Moed, n.d.).

A trend analysis provided an indication that there may be a time delay between the date a nationally oriented journal enters Scopus and the date at which its internationalization becomes apparent. Although the size of its effect is one order of magnitude smaller than that of English as publication language and OA status, this time delay should be further analysed in a follow-up study.

Also the finding that journals entering Scopus with a strong national orientation tend to have in their internationalization process an open access advantage, while more internationally oriented journals do not show such advantage, awaits further research. To identify Open Access journals, the only piece of information used in the current paper is whether or not it is included in the DOAJ/ROAD, based on information from the Scopus Source Title List. More informative and accurate information on access modality is available that could be used to analyse the bibliometric dataset available in the current study. Using all information available at the site UNPAYWALL (http://unpaywall.org/user-guides/research), Robinson, Costas & Van Leeuwen (n.d.) are devising a classification system of the various types of Open Access. A careful screening of available data on scientific-scholarly journals should also record reliable historical data on journals, so that the accuracy and validity of longitudinal analyses can be enhanced.

Finally, the current analysis is entirely based on Scopus. To obtain a more complete overview of the behaviour of nationally oriented journals, other literature databases should be analysed as well: Clarivate's Web of Science; Google Scholar; Digital Science's Dimensions; or Microsoft Academic.